\documentclass[aps,pre,superscriptaddress,twocolumn,twoside,a4paper]{revtex4-1}
\usepackage{graphicx}
\usepackage{amsmath}
\usepackage[utf8]{inputenc}
\usepackage{hyperref}
\graphicspath{{Images/}}

\begin{document}

\title{Twin Cheshire Photons} 
\author{Issam Ibnouhsein} \email{issam.ibnouhsein@cea.fr} \affiliation{CEA-Saclay/SPEC/LARSIM, 91191 Gif-sur-Yvette, France} \affiliation{Université Paris-Sud, 91405 Orsay, France} 
\author{Alexei Grinbaum} \affiliation{CEA-Saclay/SPEC/LARSIM, 91191 Gif-sur-Yvette, France}
\date{\today}

\begin{abstract}
In \cite{aharonov_quantum_2013} Aharonov {\em et al.} showed that in an experiment with both pre- and post-selection one can find a photon (the Cheshire cat) in one place and its polarization (the smile) in another, and asked whether more than two degrees of freedom could be separated in the same way. We show that this is possible and that the separation of properties from objects that carry them is in some situations even stronger. 
\end{abstract}

\maketitle

\section{Introduction}

It is usually assumed that properties are related to objects that carry them, for example slowness belongs to the turtle. Yet in his novel {\em Alice's Adventures in Wonderland} Lewis Carroll mischievously mentioned a smile with no cat. Aharonov {\em et al.} showed that quantum mechanics indeed allows such situations to happen in experiments with pre- and post-selection on both path and polarization degrees of freedom of a photon \cite{aharonov_quantum_2013}. We generalize this result to four degrees of freedom by completely dissociating the positions and the polarizations of an entangled pair of photons. In section \ref{section2}, we retrieve the `one smile no cat' result. In section \ref{section3}, we show that the disembodiment of physical properties can be even stronger, since in some situations polarizations cannot be traced back to any definite photon.

\section{One smile with no cat}
\label{section2}

Bell states are defined as follows:
\begin{align*}
&|\Phi^+\rangle=\frac{|HH\rangle+|VV\rangle}{\sqrt{2}},\\
&|\Phi^-\rangle=\frac{|HH\rangle-|VV\rangle}{\sqrt{2}},\\
&|\Psi^+\rangle=\frac{|HV\rangle+|VH\rangle}{\sqrt{2}},\\
&|\Psi^-\rangle=\frac{|HV\rangle-|VH\rangle}{\sqrt{2}},
\end{align*}
where $|H\rangle$ and $|V\rangle$ denote horizontally and vertically polarized light respectively.

\begin{figure}[!htbp]
\begin{center}
\includegraphics[scale=0.2]{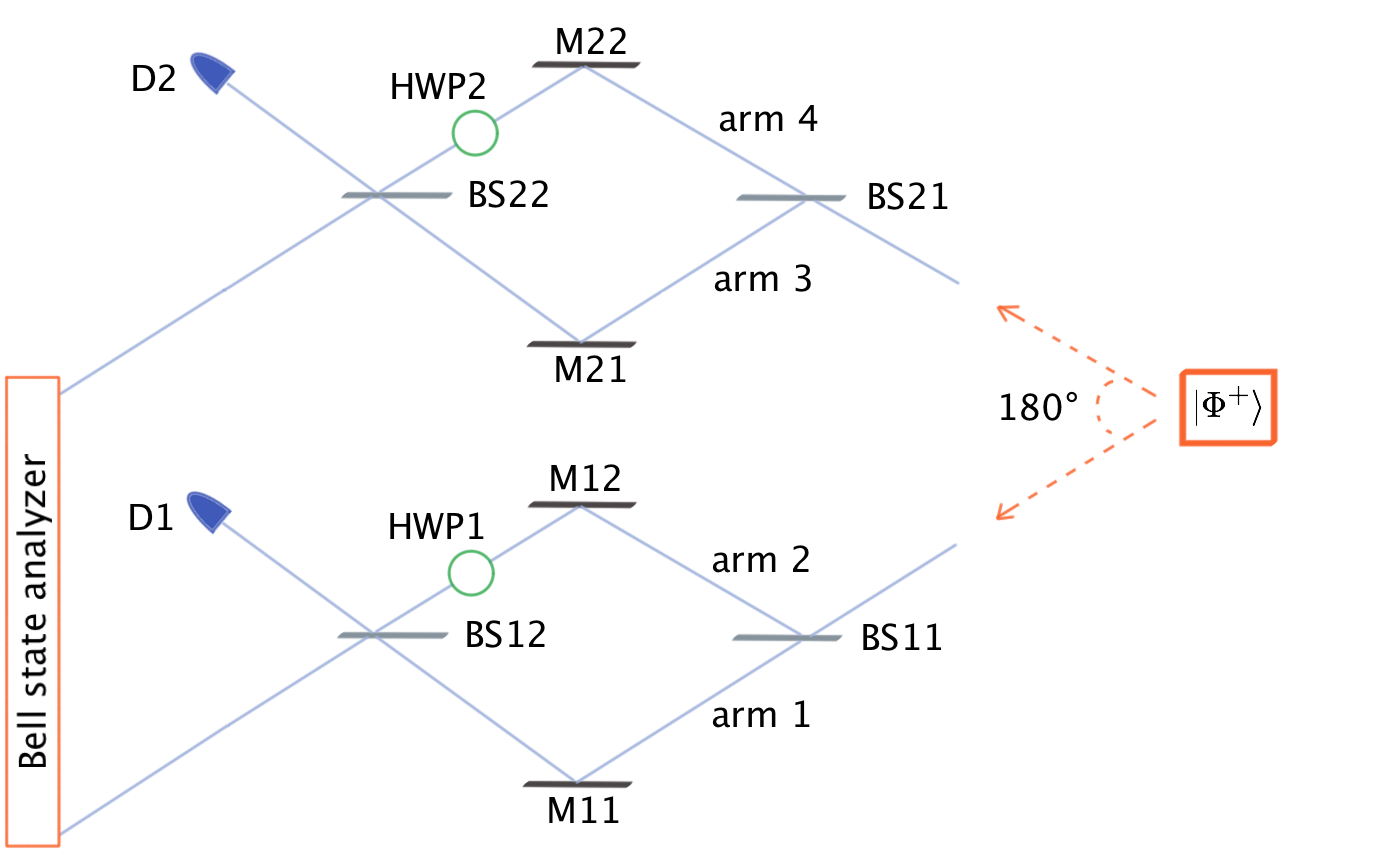}
\end{center}
\caption{An entangled pair enters two Mach-Zender interferometers from the right. The number of interferometers helps to maintain the distinction between the two photons. Half-wave plates HWP1 and HWP2 are instrumental for post-selecting $|\beta_1\rangle$.}
\label{schema}
\end{figure}

Consider a pair of entangled photons in the state $|\Phi^+\rangle$, each photon entering one interferometer as depicted in Fig.\ref{schema}. 

The state of the pair after going through beamsplitters BS11 and BS21 is the following:
\begin{equation}
|\alpha\rangle = \frac{|13\rangle + |14\rangle + |23\rangle + |24\rangle}{2} \otimes |\Phi^+\rangle,
\end{equation}
where $|ij\rangle, i=1,2, j=3,4$ is a two photons state describing the situation where the ``lower" photon goes through arm $i$ and the ``upper" one through arm $j$.

\bigskip

We seek to post-select \cite{aharonov_how_1988} the state $|\beta_1\rangle$: 
\begin{equation}
|\beta_1\rangle = \frac{1}{2}\left[\left(|13\rangle+|24\rangle \right) \otimes |\Phi^+\rangle + \left(|14\rangle -|23\rangle \right) \otimes |\Psi^-\rangle\right].
\end{equation}

Such a post-selection could be implemented experimentally using an entangled pair, two Mach-Zender interferometers, two standard photon detectors (D1 and D2), a Bell state analyzer (BSA), and two half-wave plates (HWP) which transform polarizations according to the rule:
\begin{align*}
|H\rangle &\longrightarrow |V\rangle,\\
|V\rangle &\longrightarrow -|H\rangle.
\end{align*} 
Since we consider a pair entangled only with respect to the polarization degrees of freedom, some non-linearity needs to be introduced in BSA \cite{kwiat_embedded_1998,lutkenhaus_bell_1999,barrett_symmetry_2005,barbieri_complete_2007}.

We can tune the two interferometers in such a way that, in the absence of HWPs along arms 2 and 4, if $\frac{1}{2}\left(|13\rangle + |14\rangle + |23\rangle + |24\rangle\right)$ hits BS12 and BS22, D1 and D2 won't click, \emph{i.e.} each photon emerges to the left of the corresponding interferometer. Now consider only instances in which BSA detected a $|\Phi^+\rangle$ Bell state. This ensures that the state immediatly prior to detection was $|\alpha\rangle$. The presence of a half-wave plate inside each interferometer implies that the state immediately prior to HWP1 and HWP2 was $|\beta_1\rangle$.

\bigskip

Given a pre-selection $|\alpha\rangle$ and a post-selection $|\beta_1\rangle$, the photons of the entangled pair seem to go \emph{with certainty} through arms (1,3) or (2,4). To show this, imagine that we test the presence of a photon by inserting photon detectors in the arms of the interferometers. These can be taken as non-demolition measurements in the sense that they do not absorb the photon and do not alter its polarization \cite{aharonov_quantum_2013}.
Mathematically these are defined as:
\begin{align}
\Pi_{ij} &= |ij\rangle \langle ij|
\end{align}
\begin{equation}
\Pi_{i\cdot} = \Pi_{i3} + \Pi_{i4}\,, 
\end{equation}
\begin{equation}
\Pi_{\cdot j} = \Pi_{1j} + \Pi_{2j}\,, 
\end{equation}
where $i=1,2$ (resp. $j=3,4$) denotes the arm of the lower (upper) interferometer on which a measurement is performed. The notation $i\cdot$ ($\cdot j$) represents
a measurement carried on both arms of the upper (lower) interferometer, which is equivalent to tracing out the upper (lower) particle of the entangled pair.

Suppose we insert $\Pi_{13}$. We can find 0, 1 or 2 photons. If we find one photon, then $|\alpha\rangle$ projects only on terms that contain $|\Phi^+\rangle$  and $|\beta_1\rangle$ only on terms that contain $|\Psi^-\rangle$:
$$|\alpha\rangle \longrightarrow \frac{1}{\sqrt{2}}(|14\rangle+e^{i\theta_1}|23\rangle)\otimes |\Phi^+\rangle,$$
$$|\beta_1\rangle \longrightarrow \frac{1}{\sqrt{2}}(|14\rangle-e^{i\theta_2}|23\rangle)\otimes |\Psi^-\rangle,$$ where $e^{i\theta_1}$ and $e^{i\theta_2}$ are random phase shifts due to photon number measurements. Since $\langle \Phi^+|\Psi^-\rangle = 0$, these two states are orthogonal and post-selection cannot succeed. Hence non-demolition measurements yield either 0 or 2 photons in arms 1 and 3, \emph{i.e.} the photon pair has traveled through the interferometers by way of arms (1,3) or (2,4). The same conclusion can be reached using detectors $\Pi_{24},\Pi_{14},\Pi_{23}$ or $\Pi_{1\cdot},\Pi_{2\cdot},\Pi_{\cdot 3},\Pi_{\cdot 4}$.
 
 \bigskip
Suppose the photons traveled through arms (1,3) or (2,4). Can their polarizations be found elsewhere? Let us place  polarization detectors on arms (1,4) and (2,3). Intuitively, if the polarization detector on arm 1 clicks, the polarization detector on arm 3 should also click whereas the polarization detectors on arms 2 and 4 should not: we \emph{``know"} that if one photon is in arm 1 the other photon is in arm 3. Quantum mechanics contradicts this intuition: angular momentum can be found on both arms of the pairs (1,4) or (2,3), as if one of the two photons could travel through one arm of the interferometer while its polarization traveled through another.

Formally, consider a $\frac{\pi}{4}$-rotated basis $\{|+\rangle,|-\rangle\}$ where $|+\rangle = \frac{1}{\sqrt{2}}\left(|H\rangle + |V\rangle\right)$ and $|-\rangle= \frac{1}{\sqrt{2}}\left(|H\rangle - |V\rangle\right)$. Define polarization measurement operators:
\begin{equation}
\begin{aligned}
\sigma_{zz} &= \sigma_z \otimes \sigma_z \\
& =  \left( \begin{array}{cccc} 1 & 0 & 0 & 0  \\ 0 & -1 & 0 & 0 \\  0 & 0 & -1 & 0 \\ 0 & 0 & 0 & 1 \end{array} \right) \\
&=|++\rangle \langle ++|\quad -\quad|+-\rangle \langle +-|\\
&- |-+\rangle \langle -+|\quad +\quad|--\rangle \langle --|.
\end{aligned}
\end{equation}
A measurement of polarization in arms $i$ and $j$ ($i=1,2$; $j=3,4$) corresponds to the operator:
\begin{equation}
\sigma_{zz}^{ij} = \Pi_{ij}\sigma_{zz}\,.
\end{equation}
Tracing out one of the particles corresponds to:
\begin{equation}
\sigma_{zz}^{i\cdot} = \Pi_{i\cdot}\sigma_{zz}\,,
\end{equation}
\begin{equation}
\sigma_{zz}^{\cdot j} = \Pi_{\cdot j}\sigma_{zz}\,.
\end{equation}
The observable $\sigma_{zz}^{ij}$ has three eigenvalues: $+1, -1$ and 0, corresponding to the subspaces spanned by $\{|ij\rangle\otimes|++\rangle,|ij\rangle \otimes|--\rangle\}$,$\{|ij\rangle\otimes|+-\rangle,|ij\rangle\otimes|-+\rangle\}$, and $\{|kl\rangle\otimes|++|\rangle,|kl\rangle\otimes|--\rangle,|kl\rangle\otimes|+-\rangle,|kl\rangle\otimes|-+\rangle\}_{k\neq i, l\neq j}$ respectively. The conditional probability of $\sigma_{zz}^{14}$ and $\sigma_{zz}^{23}$ yielding result +1, given that BSA detects $|\Phi^+\rangle$, is non-zero. Similarly there is a non-zero conditional probability that the measurement yields the result $-1$. Hence it seems that we found a smile with no cat, \emph{i.e.} a polarization with no photon! 

This situation is an example of the many ``paradoxes" that arise with counterfactual reasoning in quantum mechanics, as illustrated by Hardy's thought experiment \cite{hardy_quantum_1992,hardy_nonlocality_1993}. 
Simultaneous measurements of $\Pi_{ij}$ and $\sigma_{zz}^{ij}$ would always agree on the arm in which they detect the photon's position and polarization. However our measurements were not simultaneous and we had to resort to counterfactual arguments, which are contradicted by the quantum formalism even if they conform to classical intuition.

\bigskip

What happens if we consider measurements with a negligibly small disturbance, \emph{i.e.} weak measurements \cite{aharonov_revisiting_2002}?
Define the weak value of an operator $\hat A$ with pre- and post-selected states $|\psi\rangle$ and $|\phi\rangle$ as: 
\begin{equation}
{\langle \hat A \rangle}_w=\frac {\langle \phi|\hat A|\psi\rangle}{\langle \phi|\psi\rangle}\,.
\end{equation}
Calculations (see Appendix) show that:
\begin{equation}
\label{pos11}
{\langle \Pi_{13} \rangle}_w = {\langle \Pi_{24} \rangle}_w = \frac{1}{2}\,,
\end{equation}
\begin{equation}
{\langle \Pi_{14} \rangle}_w = {\langle \Pi_{23} \rangle}_w = 0\,,
\end{equation}
\begin{equation}
\label{pos13}
{\langle \Pi_{1\cdot} \rangle}_w = {\langle \Pi_{2\cdot} \rangle}_w = {\langle \Pi_{\cdot 3} \rangle}_w ={\langle \Pi_{\cdot 4} \rangle}_w = \frac{1}{2}\,,
\end{equation}
\begin{equation}
\langle\Pi_{13} + \Pi_{24}\rangle_w = \langle\Pi_{13}\rangle_w + \langle\Pi_{24}\rangle_w = 1\,.
\end{equation}
In the light of equation \eqref{pos13}, equation \eqref{pos11} can be interpreted as saying that the photon pair went with equal probability through arms (1,3) or (2,4).
Moreover only one photon travelled through each interferometer:
\begin{equation}
\label{pos15}
\langle\Pi_{1 \cdot} + \Pi_{2 \cdot}\rangle_w = \langle\Pi_{\cdot 3}\rangle_w + \langle\Pi_{\cdot 4}\rangle_w = 1.
\end{equation}

Similarly for weak values of the polarization measurement operators (see Appendix):
\begin{equation}
\label{sig61}
\langle\sigma_{zz}^{13}\rangle_w = \langle\sigma_{zz}^{24}\rangle_w = \frac{1}{2}\,,
\end{equation}
\begin{equation}
\label{eq.1}
\langle\sigma_{zz}^{14}\rangle_w = \langle\sigma_{zz}^{23}\rangle_w = 0\,,
\end{equation}
\begin{equation}
\langle\sigma_{zz}^{1\cdot}\rangle_w = \langle\sigma_{zz}^{2\cdot}\rangle_w = \langle\sigma_{zz}^{\cdot 3}\rangle_w = \langle\sigma_{zz}^{\cdot 4}\rangle_w = \frac{1}{2}\,.
\end{equation}
If we make a simultaneous weak measurement of $\sigma_{zz}^{13}$ and $ \sigma_{zz}^{24}$, we get the total weak value:
\begin{equation}
\langle\sigma_{zz}^{13} + \sigma_{zz}^{24}\rangle_w = \langle\sigma_{zz}^{13}\rangle_w + \langle\sigma_{zz}^{24}\rangle_w = 1\,,
\end{equation}
which together with equation \eqref{eq.1} says that we have found polarizations traveling along arms (1,3) or (2,4) with equal probability. Furthermore it is true that: 
\begin{equation}
\label{sig201}
\langle\sigma_{zz}^{1 \cdot} + \sigma_{zz}^{2 \cdot}\rangle_w = \langle\sigma_{zz}^{\cdot 3}\rangle_w +\langle\sigma_{zz}^{\cdot 4}\rangle_w = 1\,,
\end{equation}
which is consistent with the fact that there is one polarization in each interferometer. No Cheshire cat until now!

Interesting effects for an entangled pair of photons appear only when weak polarization measurements are carried along orthogonal directions. Consider the following polarization measurement operator:
\begin{equation}
\begin{aligned}
\sigma_{zx} &= \sigma_z \otimes \sigma_x \\ 
& =  \left( \begin{array}{cccc} 0 & 1 & 0 & 0  \\ 1 & 0 & 0 & 0 \\  0 & 0 & 0 & -1 \\ 0 & 0 & -1 & 0 \end{array} \right) \\
&= |+-\rangle \langle ++| \quad + \quad |++\rangle \langle +-| \\
&-|--\rangle \langle -+| \quad - \quad |-+\rangle \langle --|\,.
\end{aligned}
\end{equation}
Weak values for the different $\sigma_{zx}^{ij}$ operators are (see Appendix):
\begin{equation}
\label{sigzx1}
\langle\sigma_{zx}^{13}\rangle_w = \langle\sigma_{zx}^{24}\rangle_w = 0\,, 
\end{equation}
\begin{equation}
\langle\sigma_{zx}^{14}\rangle_w = \langle\sigma_{zx}^{23}\rangle_w = \frac{1}{2}\,, 
\end{equation}
\begin{equation}
\label{sigzx2}
\langle\sigma_{zx}^{1\cdot}\rangle_w = \langle\sigma_{zx}^{2\cdot}\rangle_w = \langle\sigma_{zx}^{\cdot 3}\rangle_w = \langle\sigma_{zx}^{\cdot 4}\rangle_w = \frac{1}{2}\,. 
\end{equation}
Athough the photons of the pair go only through arms (1,3) or (2,4) with equal probability:
\begin{equation*}
{\langle \Pi_{13} \rangle}_w = {\langle \Pi_{24} \rangle}_w = \frac{1}{2}\,,
\end{equation*}
\begin{equation*}
\langle\Pi_{14}\rangle_w = \langle\Pi_{23}\rangle_w = 0\,,
\end{equation*} 
polarizations do not: 
\begin{equation*}
\langle\sigma_{zx}^{13}\rangle_w = \langle\sigma_{zx}^{24}\rangle_w = 0\,.
\end{equation*}
Indeed, polarization travel along arms (1,4) or (2,3) with equal probability:
\begin{equation*}
\langle\sigma_{zx}^{14}\rangle_w = \langle\sigma_{zx}^{23}\rangle_w = \frac{1}{2}\,.
\end{equation*}
Hence the photon's position and its polarization traveled along different arms in one of the two interferometers, chosen at random. The crucial point is that \emph{all} these measurement are performed \emph{simultaneously}. Therefore we have found a quantum Cheshire cat \cite{aharonov_quantum_2013}.

\section{Two smiles with no cats}
\label{section3}

In section \ref{section2}, a photon's polarization traveled along a different arm than its position. Now we shall separate all polarizations from all photons positions.

Consider a post-selected state:
\begin{equation}
|\beta_2\rangle = \frac{1}{2}\left[|13\rangle \otimes |\Phi^+\rangle + (|24\rangle - |14\rangle+|23\rangle) \otimes |\Psi^-\rangle\right].
\end{equation}
This post-selection constrains the photons of the pair to travel through arms 1 and 3, one photon entering each interferometer.

As before (see Appendix):
\begin{equation}
\label{pos3}
{\langle \Pi_{13} \rangle}_w = 1\,, 
\end{equation}
\begin{equation}
{\langle \Pi_{24} \rangle}_w = {\langle \Pi_{14} \rangle}_w = {\langle \Pi_{23} \rangle}_w = 0\,, 
\end{equation}
\begin{equation}
{\langle \Pi_{1\cdot} \rangle}_w =  {\langle \Pi_{\cdot 3} \rangle}_w = 1\,, 
\end{equation}
\begin{equation}
\label{pos29}
{\langle \Pi_{2\cdot} \rangle}_w = {\langle \Pi_{\cdot 4} \rangle}_w = 0\,. 
\end{equation}
Equation \eqref{pos3} confirms that the photons of the pair went through arms 1 and 3.
Measuring polarizations along the same direction for both photons yields no interesting results:
\begin{equation}
\label{sig13}
\langle\sigma_{zz}^{13}\rangle_w = 1\,, 
\end{equation}
\begin{equation}
\label{sig14}
\langle\sigma_{zz}^{24}\rangle_w = \langle\sigma_{zz}^{14}\rangle_w = \langle\sigma_{zz}^{23}\rangle_w = 0\,,
\end{equation}
\begin{equation}
\label{sig15}
\langle\sigma_{zz}^{1\cdot}\rangle_w =  \langle\sigma_{zz}^{\cdot 3}\rangle_w = 1\,,
\end{equation}
\begin{equation}
\label{sig16}
\langle\sigma_{zz}^{2\cdot}\rangle_w = \langle\sigma_{zz}^{\cdot 4}\rangle_w = 0\,. 
\end{equation}
If measurements are carried out along orthogonal directions, then (see Appendix):
\begin{equation}
\label{sig17}
\langle\sigma_{zx}^{13}\rangle_w = 0\,,
\end{equation}
\begin{equation}
\label{sig18}
\langle\sigma_{zx}^{24}\rangle_w = 1\,,
\end{equation}
\begin{equation}
\label{sig19}
\langle\sigma_{zx}^{14}\rangle_w = -\langle\sigma_{zx}^{23}\rangle_w = -1\,, 
\end{equation}
\begin{equation}
\label{sig110}
\langle\sigma_{zx}^{1\cdot}\rangle_w = -\langle\sigma_{zx}^{\cdot 3}\rangle_w = -1\,, 
\end{equation}
\begin{equation}
\label{sig111}
\langle\sigma_{zx}^{2\cdot}\rangle_w = 2\,,
\end{equation}
\begin{equation}
\label{sig112}
\langle\sigma_{zx}^{\cdot 4}\rangle_w = 0\,. 
\end{equation}
Hence a conclusion: though the photons of the pair went with certainty through arms 1 and 3 of the interferometers, as confirmed by \eqref{pos3}, no polarization pair traveled through these arms, as shown by \eqref{sig17}. The polarization pair traveled along arms (2,3) and (2,4) with certainty, as shown by \eqref{sig18} and \eqref{sig19}, and quantum mechanics compensates for the excess in the number of pairs by letting -1 polarization pair travel through arms (1,4)! This is consistent with the fact that there is one photon in each interferometer since only one polarization travels through each interferometer: 
\begin{equation}
\label{sig113}
\langle\sigma_{zx}^{1\cdot}\rangle_w + \langle\sigma_{zx}^{2\cdot}\rangle_w = 1, \hspace{0.2cm}
\langle\sigma_{zx}^{\cdot 3}\rangle_w + \langle\sigma_{zx}^{\cdot 4}\rangle_w = 1\,.
\end{equation}
Note that $\langle\sigma_{zx}^{2\cdot}\rangle_w = 2$, meaning that any system weakly interacting with arm 2 would feel the presence of two polarizations whereas only one photon entered the interferometer, in agreement with the interpretation of Aharonov \emph{et al.} \cite{aharonov_how_1988}. Still only one polarization traveled through this interferometer since a measurement on arm 1 would detect ``-1" polarization, bringing the total to 1. Here the question ``which photon polarization?'' makes no sense, unlike the Cheshire cat experiment in section \ref{section2}.
\bigskip

\section{Conclusions}

We showed that the disembodiment of physical properties from objects to which they supposedly belong in pre- and post-selected experiments can be generalized to more than two degrees of freedom: two-photon positions and polarizations. The dissociation is even stronger since, in some cases, the correspondence between objects and properties is completely lost.

Our results concerned the polarisation of entangled photons, but they can be extended to more general situations, \emph{e.g.} separation of the spin from the charge of an electron, or that of the mass of an atom from its position \cite{aharonov_quantum_2013}. This could help to realize precision measurements by spatially separating various properties of the measured system. An interesting task would be to retrieve these results in the two-state vector formalism \cite{aharonov_two-state_2001} and to understand how the behaviour of the polarization of each state vector can give rise to the ``paradoxical" effects due to post-selection.
 
\bibliographystyle{apsrev4-1}
\bibliography{cheshire_twins.bib}

\newpage
\section*{Appendix}
A computation gives: $\langle \beta_1|\alpha\rangle = \frac{1}{2}$ and $\langle \beta_2|\alpha\rangle = \frac{1}{4}$.

\bigskip

The weak value of $\Pi_{ij}$ is calculated as follows:
\begin{align*}
\Pi_{ij}|\alpha\rangle &=|ij\rangle \langle ij| \left[\frac{|13\rangle + |14\rangle + |23\rangle + |24\rangle}{2}\right] \otimes |\Phi^+\rangle\\
 &= \frac{|ij\rangle}{2}|\Phi^+\rangle.
\end{align*}
A projection of $\Pi_{ij} |\alpha\rangle$ on $\langle \beta_1|$ or $\langle \beta_2|$ will depend on $i$ and $j$, leading to equations \eqref{pos11} to \eqref{pos15} and \eqref{pos3} to \eqref{pos29}.

Similarly, the weak value of $\sigma_{zz}^{ij}$ is calculated as follows:
\begin{align*}
\sigma_{zz}^{ij}|\alpha\rangle &= \Pi_{ij} \sigma_{zz} |\alpha\rangle \\
&=|ij\rangle \langle ij|\left(|++\rangle \langle ++|-|+-\rangle \langle +-| \right. \\
& \left. -|-+\rangle \langle -+|+|--\rangle \langle --|\right) |\alpha\rangle\\ .
\end{align*}
As an intermediate step, compute the scalar products:
\begin{align*}
&\langle ++|\Phi^+\rangle =\langle --|\Phi^+\rangle =\frac{1}{\sqrt{2}},\\
&\langle +-|\Phi^+\rangle =\langle -+|\Phi^+\rangle = 0.\\
\end{align*}
Inserting these values we get:
\begin{align*}
\sigma_{zz}^{ij} |\alpha\rangle&=\frac{|ij\rangle}{2\sqrt{2}}\left(|++\rangle+|--\rangle\right)\\
&=\frac{|ij\rangle}{2}|\Phi^+\rangle.
\end{align*}
A further projection on $\langle \beta_1|$ or $\langle \beta_2|$ will depend on $i$ and $j$. Hence the results of equations \eqref{sig61} to \eqref{sig201} and \eqref{sig13} to \eqref{sig16}.

\bigskip

Similarly, the weak value of $\sigma_{zx}^{ij}$ can be computed as follows:
\begin{align*}
\sigma_{zx}^{ij}|\alpha\rangle &= \Pi_{ij} \sigma_{zx} |\alpha\rangle \\
&= |ij\rangle\langle ij| \left(|+-\rangle \langle ++|+|++\rangle \langle +-|\right.\\
&\left. -|--\rangle \langle -+|-|-+\rangle \langle --|\right)|\alpha\rangle \\ 
&=\frac{|ij\rangle}{2\sqrt{2}}\left(|+-\rangle-|-+\rangle\right)\\
&=\frac{|ij\rangle}{2}|\Psi^-\rangle.
\end{align*}
A further projection on $\langle \beta_1|$ or $\langle \beta_2|$ will depend on $i$ and $j$. Hence the results of equations \eqref{sigzx1} to \eqref{sigzx2} and \eqref{sig17} to \eqref{sig113}.

\end{document}